\documentclass{article}
\usepackage{wrapfig,epsfig,epsf}
\usepackage{graphics}
\usepackage{amsmath}
\begin{document}
\title{ 
\large \flushleft
Preprint 9, 2014\\
P.N.~Lebedev Physical Institute of the Russian Academy of Sciences\\[20pt]
\LARGE \bf \center {The FAKERAT Software in the International Interferometric Project
RADIOASTRON with Very Long Space-Ground  Baselines}}
\author{ \Large\bf V.I.~Zhuravlev\\[10pt]
\it Astro Space Center, Lebedev Physical Institute, Russian Academy of
Sciences, \\
Profsoyznaya str. 84/32, Moscow 117997, Russia\\
zhur@asc.rssi.ru}
\date{March 2014}
\maketitle
\begin{abstract}
We present the description of the FAKERAT software developed for planning 
Very Long Baseline Interferometry observations from space (space-VLBI). 
The results of the planned observations using the FAKERAT package 
during the first two years after 
launch of the space radio telescope (SRT) in the space-ground 
interferometer modes are reported.
\end{abstract}
\section{Introduction}
During the first two years the space radio observatory was operated 
according to the Early Science Program (ESP) of the RadioAstron 
project \cite{ak, ka}. During that stage, the FAKERAT software was tested 
along with testing the on-board scientific complex. This software may 
be of interest to potential users who wishe to simulate RadioAstron 
observation, taking into account structural constraints on 
spacecraft (SC) orientation. Science observations with the RadioAstron 
SRT are possible only when communication with the ground tracking  
station\footnote{In order to transfer data
in the interferometric mode during ESP stage, only one highly-informative
radio channel (HIRC) of the tracking station prepared in Puschino based on
RT-22 radio telescope was used. When this paper was written the second 
HIRC was prepared on 43 m equatorial radio telescope in NRAO.} 
for data transfer and frequency synchronization is in operation.
These constraints  may make it impossible for SRT to observe a given 
source under certain conditions.
As a result of simulating, the information about the possibility of carrying 
out interferometric sessions for each particular object may be taken. 
The simulation takes into account the detailes of the defined 
scientific task, observation duration, observation date, wavelength 
range, projection of the interferometer baseline and (u,v)-coverage.

FAKERAT is based on the programs included into "Caltech VLBI Analysis 
Program" package developed by T.J. Pearson (Caltech) in 1979 in order 
to support planning and analysis of interferometric observations in 
experiments with ground baselines. In 1983 D.L. Meier (NASA Jet 
Propulsion Laboratory (JPL)), added the ability to simulate space 
VLBI experiments as part of the QUASAT mission development. The 
software was further developed by D.W. Murphy (JPL) \cite{mu1, mu2}. 
He introduced functional constraints for the space radio telescope 
VSOP specifically designed for interferometry and added graphical 
interface. Graphical interface significantly simplified the work 
related to the experiment planning and analysis of scientific 
perspectives on interferometric observations. During that stage 
of software development the FAKESAT package was created. The 
software was written in FORTRAN and runs on both SUN and HP 
workstations. It should be relatively easy implemented for UNIX platforms.

In spring of 2011, half a year before RadioAstron launch, we modified 
FAKESAT taking into account necessary constraints and other peculiarities 
of the RadioAstron project. Modification included the full replacement of 
the orbital block. In new orbital block we considered perturbation of 
orbital elements, introduced new functional constraints specified by 
orientation SC, the ground tracking and scientific data acquisition 
station and changed graphical interface to make software more easy to use.

Binary and source code of the modified version  known as 
FAKERAT\footnote{The name of package
contains word "FAKE" (from FAKESAT) and also abbreviation of name of
RadioAsTron project. Alternatively, it literally means (from English) 
"fake rat", i.o. an unreal creature object in the Universe to find
out the truth.} is available on the ASC FIAN 
website\footnote{http://www.asc.rssi.ru/radioastron/software/soft.html}. 
Installation and usage instructions can be found 
in RadioAstron User Handbook\footnote{RadioAstron User Handbook, 2012:
http://www.asc.rssi.ru/radioastron/documents/en/rauh.pdf}.

\section{Orbital Motion of the Spacecraft}

	In FAKERAT, the orbital motion of the spacecraft is described by 
the table of $x$, $y$, $z$ coordinate values and components of $v_x$, 
$v_y$, $v_z$ velocity vector in the geocentric coordinate system. 
Geocentric coordinates and orbital velocities as a function of time 
are calculated in Keldysh Institute of Applied Mathematics, Russian 
Academy of Sciences (IAM RAN).

	Orbital motion depends on external forces acting on 
the spacecraft. 
Not all  orbit perturbations including influence of the Sun
and Moon are defined a priori with fine 
precision. 
The deviation can also arise under the influence of the forces generated
by the equipment of SC.
For example, it appears due to reactive forces of stabilizing 
system engines when unloading gyroscopes. Therefore, approximately 
once every 2-3 months IAM RAN provides new refined tabular values of 
coordinates and velocities for SC. It is also important to specify 
orbit after its correction. Fortunately, such manipulations with 
SC should be done seldom.
\begin{figure}
\rotatebox{0}{\includegraphics[width=1.\textwidth]{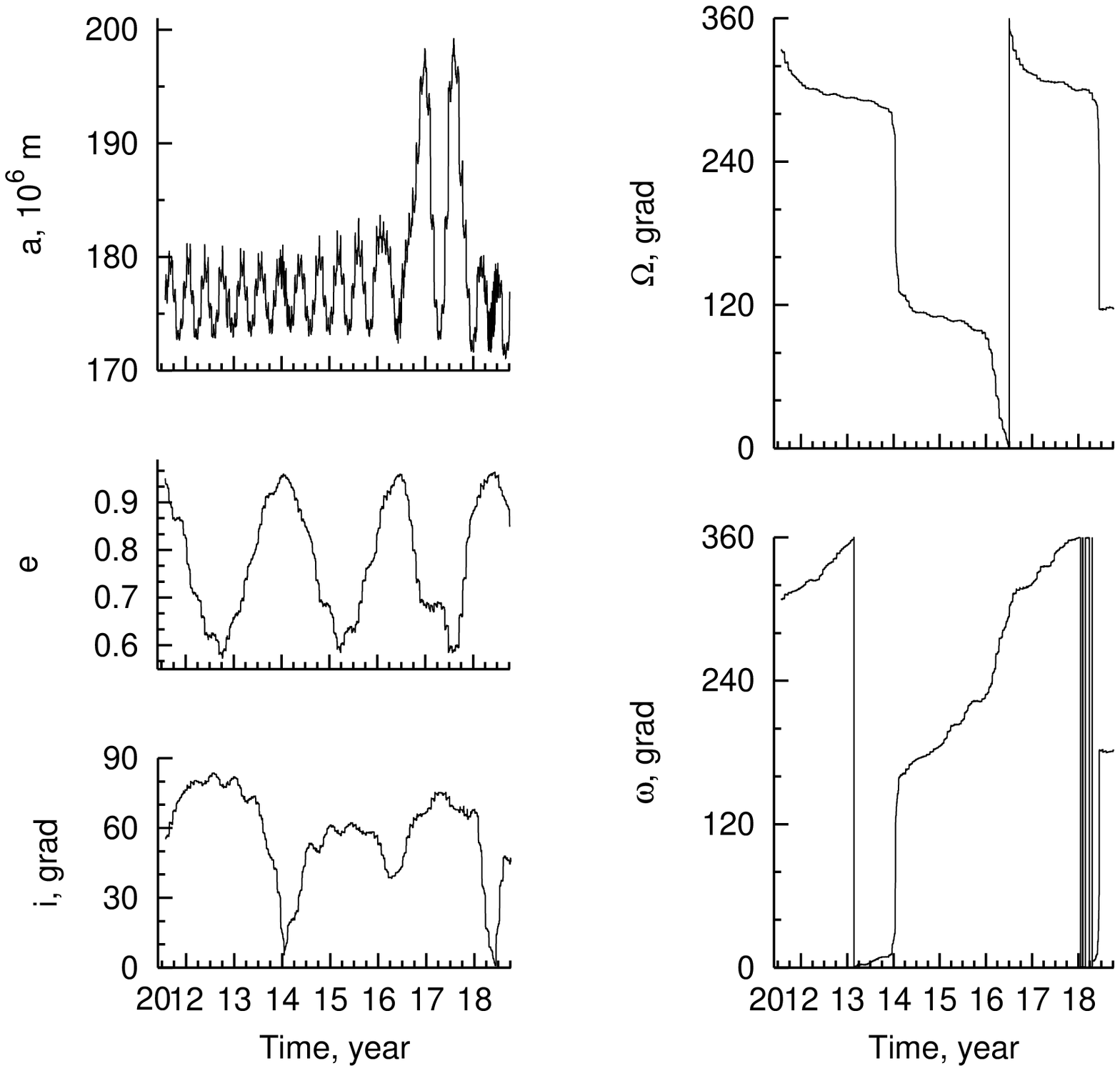}}
\caption{The variations of orbital elements in several years: 
$Á$ - semi-major axis (top, left), 
$Å$ - eccentricity (left, middle), 
$i$ - inclination angle (an angle between the orbit plane and 
the equatorial plane (bottom left), 
$\Omega$ - right ascension of the ascending node (top, right), and 
$\omega$ - angle between the ascending node and the perigee (bottom, right). } 
\label{fig:1}
\end{figure}

	In order to get response of an interferometer, it is necessary to 
know the most precise values of SC coordinates and velocities. It is not an 
easy task, and, as a rule, orbital calculation for the correlator is 
limited by the duration of the observations. Positional accuracy of reconstructed 
orbit for the data processing in the correlator must be no worse than $\pm$500 m 
and velocity accuracy no worse than  $\pm$2 cm/c in the three coordinates. 
More detailed information 
about orbit reconstruction can be found in  \cite{ak, ka}. 
Accuracy 
requirements for orbital elements in FAKERAT package are by 1.5-2 times 
less. It allows  us to make prediction of the orbit for 5-6 years ahead, 
which is especially important for planning future observations.

	The calculation of the evolution of six orbital elements until 
the middle of 2019 is shown in Fig.~\ref{fig:1}. We can see, that due to the 
impact of the perturbations the length of the major semi-axis changes from 
170000 km to  200000 km, eccentricity - from 0.57 to 0.97, and orbit inclination -- 
from 0$^\circ$.4 to 84$^\circ$.

	There are several derived parameters: a(1+e) is the orbit apocenter, 
$a(1-e)$ is the orbit pericenter, and $P$ is the rotation period. The rotation 
period with known the length of the major semi-axis is given by:
\begin{eqnarray}
P=2\pi a^{3/2}/\sqrt{\mu}
\label{q1}
\end{eqnarray}
where  $\mu= 3.986\cdot 10^5$ km$^3$/c$^2$ is a coefficient equal to the 
product of the gravitational constant and the Earth mass.

	According to the Equation~(\ref{q1}) and prediction model of the 
evolution of orbital elements given in Fig.~\ref{fig:1}, rotation period will 
vary from 8.3 to 9.2 days until 2017, and after 2017 its 
maximum value will increase up to 10.2 days.

	The table of coordinate values and components of velocity vector
is necessary for FAKERAT to know SC coordinates and velocities at 
intermediate time points relative to their tabulated values. To 
obtain such values, we approximate orbit using ellipse with the 
center of gravity of the Earth at one of its focuses. Here we took 
into consideration that values of $x$, $y$, $z$, $v_x$, $v_y$ and $y_z$
 at any point 
of the orbit identically related to the solution of Kepler's equation. 
Each new line of the table gives new improved orbital approximation. 
Major semi-axis of an ellipse, eccentricity, ellipse orientation in 
space, and also SC localization on the orbit were defined by six 
above-mentioned values. Orbital plane is defined by the orbit inclination 
and longitude of the ascending node. Pericenter is defined by the 
angular distance from the ascending node to the orbital pericenter in 
the direction of SC motion. Timing is defined by moment of passing SC 
through the pericenter. Angular distance of any point in the orbit is 
defined relative to the pericenter:
\begin{eqnarray}
u=\omega+\theta \nonumber
\end{eqnarray}
where $\theta$ is the true anomaly of this arbitrary point. And finally, 
form and size of the orbit are defined by the major semi-axis and the 
eccentricity.

\begin{figure}
\rotatebox{-90}{\includegraphics[width=0.73\textwidth]{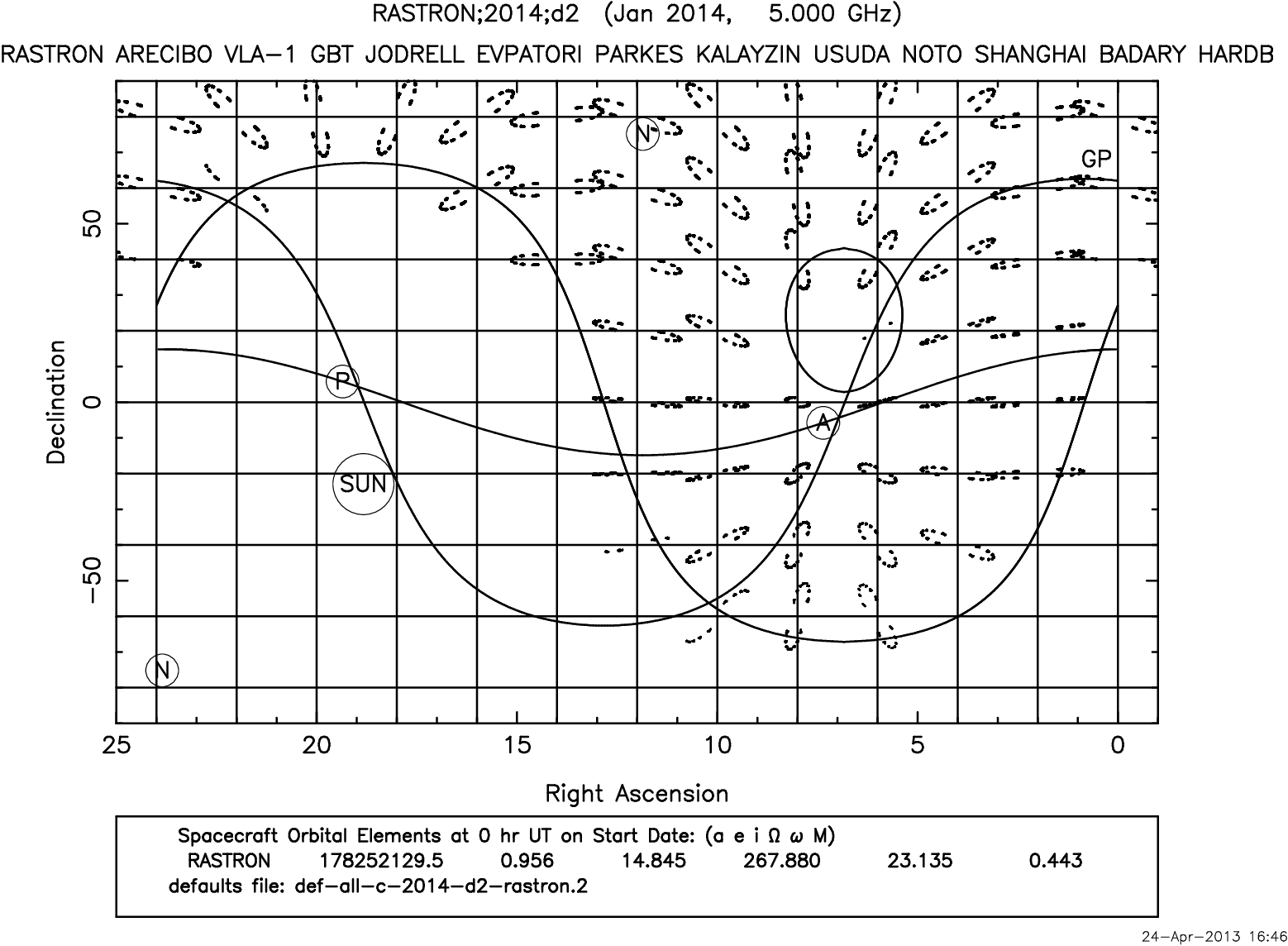}}\\
\rotatebox{-90}{\includegraphics[width=0.73\textwidth]{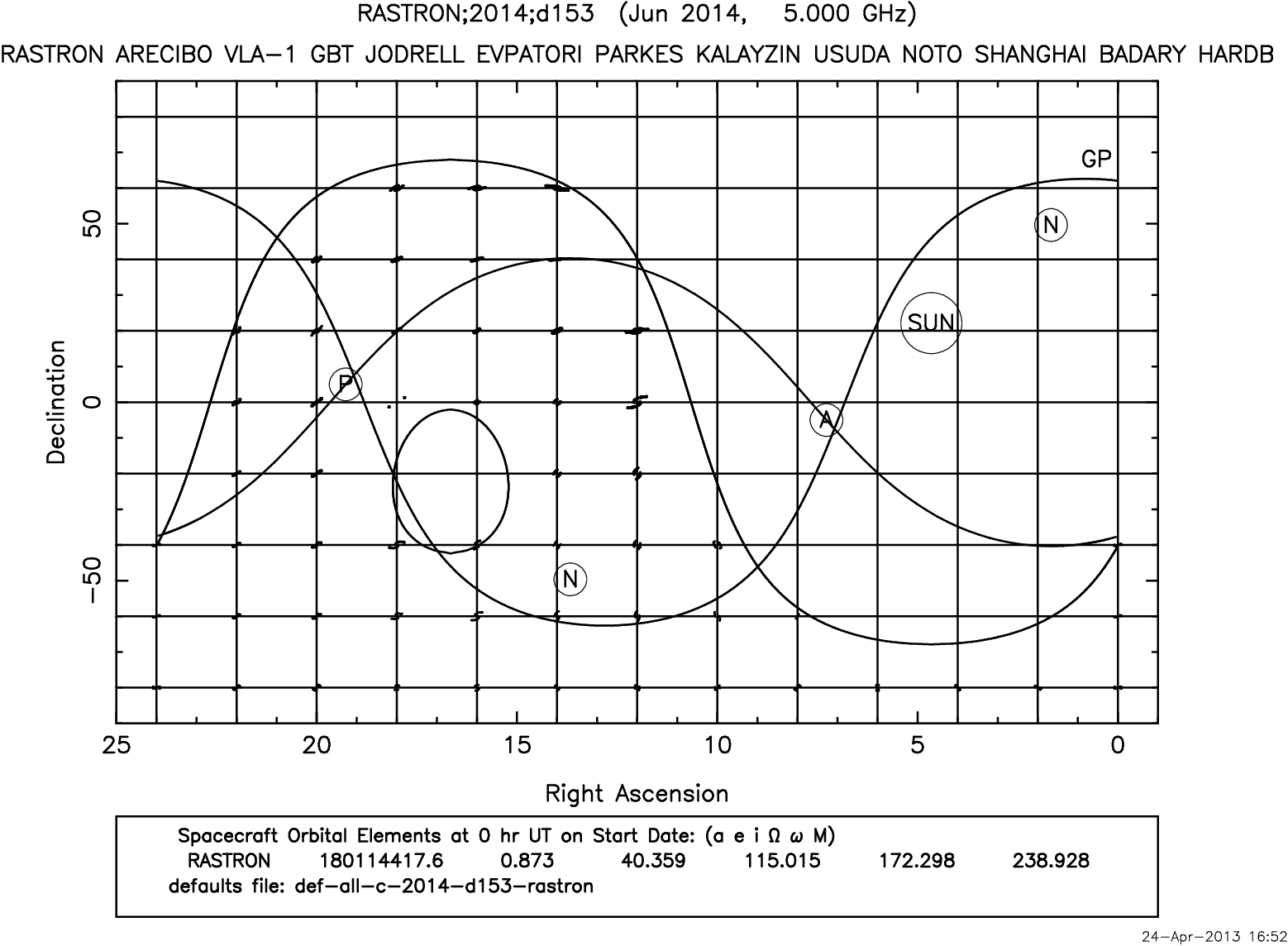}}
\caption{
An example of the all-sky (u,v)-plots at the moment favorable for 
observation when the Sun is in the lower part of the celestial sphere (top)
and at the moment when the Sun is in the upper part of the celestial 
sphere (bottom). }
\label{fig:2}
\end{figure}
	Filling (u,v)-coverage is mostly defined by the values of the 
orbital elements. High-elliptical orbit was chosen in order to get 
ultra-high angular resolution of the objects. The decrease in the 
degree of filling (u,v)-coverage was supposed to be compensated by 
orbital evolution under the influence of perturbation from Sun and 
Moon. It is worthwhile to say that maximum angular resolution is 
achieved in the direction of normal line towards the orbit plane 
(see Fig.~\ref{fig:2}). Coordinates of normal lines for the northern and 
southern celestial hemisphere are defined by the following expressions: 
($\Omega$-90$^\circ$, 90$^\circ$-$i$) and 
($\Omega$+90$^\circ$, $i$-90$^\circ$) respectively. Sources in the areas 
close to the normal line towards the plane in the southern celestial 
hemisphere are rarely observed due to the functional constraints.

	In Fig.~\ref{fig:2} normal lines towards the plane are marked as 
\begin{picture}(20,20)
\put(10,5){\circle{15}}
\put(5,1){$N$}
\end{picture}. 
In observations  of radio sources in the areas close to the normal line, 
(u,v)-tracks have elliptical structure. This structure contains big gaps. 
Due to orbit evolution the same sources my be located at one epoch close 
to the normal, and at other epoch close to the orbit plane. As a result 
gaps in (u,v)-tracks can be reduces. This reduction takes place due to 
the simultaneous degradation of angular resolution of the source. In 
extreme case, when the source is in the orbital plane, i.e. if the 
distance from the source to the normal line is 90$^\circ$, (u,v)-tracks have 
linear structure. A more detailed description of Fig.~\ref{fig:2} will be 
given below.

\section{Spacecraft Constraints}
		
	There are a number of constraints for SC which make technically 
impossible to provide observations. Those constraints are given below.

\subsection{Thermal Constraints}
	Observations are not permitted: 
\begin{itemize}
\item if the angle between SRT electric axis (X) and direction towards 
SUN center is less than 90$^\circ$; 
\item if the angle between SRT electric axis and direction towards 
SUN center is more than 165$^\circ$; 
\item if the distance from Earth center to SC is less than 20000 km 
and a radio source is located at the angular distance less than 30$^\circ$ 
from the center of Earth's disk.
\end{itemize}
\subsection{Constraints Connected with the Solar Battery Panel}
	The angle between the normal to the solar panels plane and 
the direction to the Sun must not exceed 10$^\circ$. Solar panels 
can be rotated around the rotation Y-axis only.

\subsection{Constraints Connected with the Earth and the Moon.}
 	Observations are not permitted:
\begin{itemize}
\item if the distance from the radio source to the nearest edge of 
the Earth is less than 5$^\circ$;
\item if the distance from the radio source to the center of the 
Moon's disk is less than 5$^\circ$.
\end{itemize}

\subsection{Constraints Connected with Star Sensors}
	On-board control system includes three star sensors 
(AX1, AX2 and AX3). In normal mode, only two of them are required. 
The optical axes of two star sensors (AX1) and (AX2) are placed
in the semi-plane turned around the axis parallel to the rotation axis 
of the solar battery panely (Y) at 15$^\circ$ from the semi-plane YOZ 
in the direction -X axis, and angle between the AX1 and the -OY axes is 
45$^\circ$ and the same angle exist between the AX2 and the +OY axes.
The axis of the third star sensor (AX3) is  located in the semi-plane 
XOZ with an inclination from the -X axis by 30$^\circ$ 
in the direction of the -Z axis. 

The defined above 
coordinate system is right-handed and orthogonal. 

According to the 
measurements made by the Lavochkin Scientific and Production Association 
(SPA) in July 2012, the star sensors coordinates in the SRT system 
have next directional cosines:

\begin{tabular*}{60mm}{rrrr}
	AX1: &  -0.86640913 & -0.00055799 & -0.499933447\\
	AX2: &  -0.18425467 & -0.70842875 & -0.68130677\\
	AX3: & -0.18282000  & 0.70791153  &  -0.68223024\\
\end{tabular*}

	The angle between the axes of each of the two working star 
sensors and an interfering celestial body should exceed: 
for Sun - 40$^\circ$ (from the Sun center); 
for Moon - 30$^\circ$ (from the Moon center); 
for Earth - 30$^\circ$ (from the nearest edge of Earth).

\subsection{Constraints Connected with High-Gain Communication Antenna (HGCA) 
"Board-Earth"}
HGCA provides communication with the ground tracking station (GTS) 
in order to transfer scientific and service information, and also 
enables frequency synchronization. Radio source can be observed 
only in case if such connection exists. Initial angular position 
GTS relative to the STR coordinate system is defined by the directional 
cosines:

\begin{tabular*}{60mm}{rrrr}
X axis & -0.95585102 & -0.00818895 & 0.29373758\\
Y axis & -0.00960202 & 0.99994823 & -0.00336890\\
Z axis & -0.29369479 & -0.00604064 & -0.95588015\\
\end{tabular*}

\subsubsection{Algorithm for Calculating HGCA Drive Guidance Angles}
    When calculating $\psi$ and $\vartheta$ angles, 
 anyone should take into 
account $\delta_1$ and $\delta_2$ angles of the actual electric axis 
of HGCA. Where $\delta_1$ -- angular deviation of HGCA electric 
axis from the plane XOY and $\delta_2$ -- angular deviation of HGCA 
electric axis onto the plane XOY from "OX" axis. Positive 
direction of $\delta_1$ is towards "-OZ" axis and  positive direction 
of $\delta_2$ is towards "OY". When drive angles ($\psi$ and $\vartheta$) 
are equal to zero 
 $\delta_1$=14$^\circ$.67 and $\delta_2$=0$^\circ$.

	Let us assume that $r$ is a unit radius-vector of GTS point in 
the on-board coordinate system. Then the following algorithm is 
implemented:
\begin{eqnarray}
\psi=atan2\biggl(\frac{-r_z}{\sqrt{1-r^2_y}},\frac{r_x}{\sqrt{1-r^2_y}}\biggr)-\arcsin(\frac{\sin\delta_1}{\sqrt{1-r^2_y}}\biggr) \nonumber \\
\vartheta=\arcsin\biggl(\frac{r_y}{\cos{\delta_1}}\biggr) - \delta_2 \nonumber
\end{eqnarray}
and condition according to which $\psi$ and $\vartheta$  angles belong 
to the operating range is checked: 
\begin{eqnarray}
-75^\circ \le \psi \le 90^\circ;
90^\circ \le \vartheta \le 90^\circ. \nonumber
\end{eqnarray}

The right-handed coordinate system is accepted for positive direction 
of HGCA drive rotation.

\subsection{Constraints Connected with  the Ground Tracking and 
Scientific Data Acquisition Station}

	Scientific and service information in the project RadioAstron 
is received by means of 22 meter ground radio telescope of the Puschino 
Radio Astronomical Observatory (PRAO). The radio telescope should ensure 
that SRT tracking is performed during the communication session. As it 
was mentioned above, now it is possible to use the ground tracking 
station Green Bank with the 43-m radio telescope.

	Permitted range of GTS rotation angles in Puschino was specified 
during ESP: azimuth $A$ - from 6$^\circ$ to 354$^\circ$ and 
height $h$ - from 10$^\circ$ to 84$^\circ$. 

Permitted 
range of 43 m radio telescope limited azimuth angles from 82$^\circ$.5 to
277$^\circ$.5 and the height must be greater then 11$^\circ$.

\begin{figure}
\rotatebox{-90}{\includegraphics[width=0.73\textwidth]{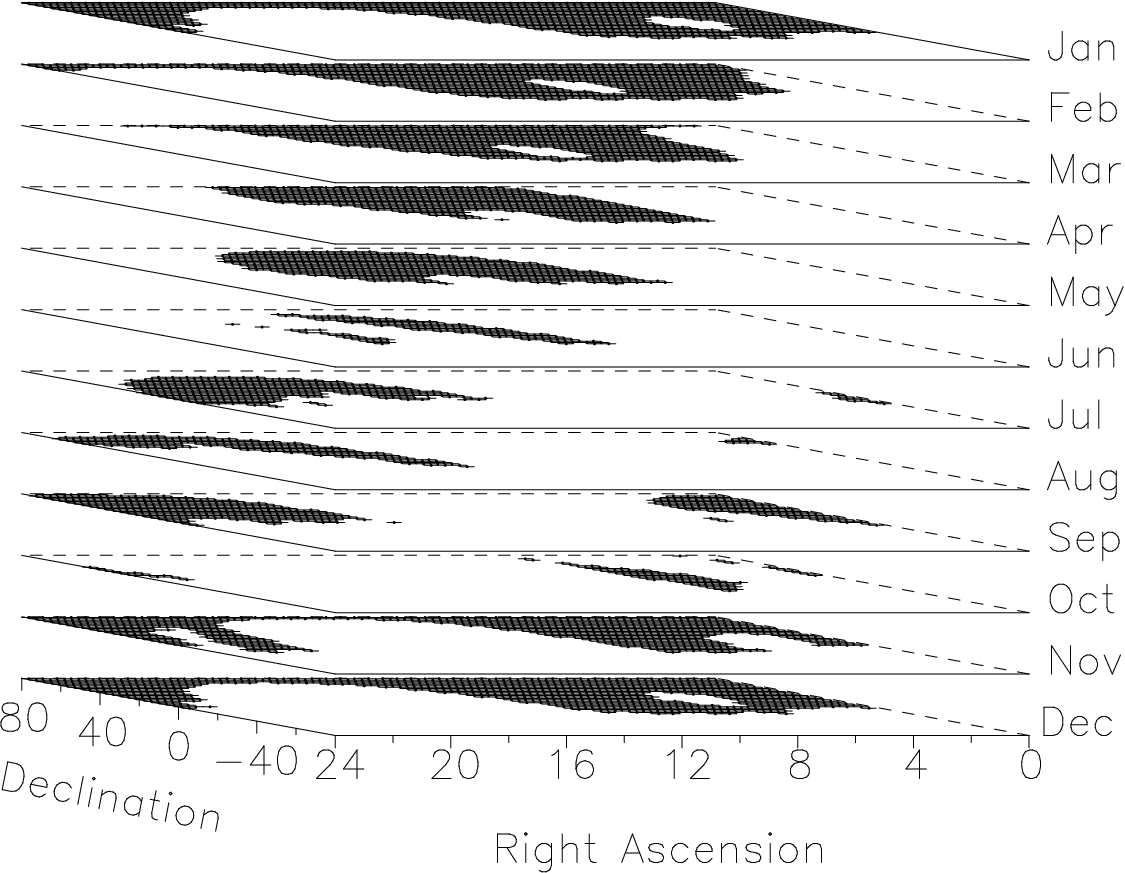}}
\caption{
Areas available for observation (dark regions) located according to 
the months within 2014.}
\label{fig:3}
\end{figure}

\section{Season Variations of Available Sky Area}

	 Fig.~\ref{fig:3} shows twelve areas for observations 
on the celestial sphere in increments of one month. In order 
to receive this image, we scanned the celestial sphere with a step in 
right ascension 30 minute and  declination -- 4$^\circ$. 
The possibility of making observations using  RadioAstron 
spacecraft and co-observing ground radio telescopes (Table~\ref{tab:1}) 
for each node was checked.  Necessary integration time was at 
least one hour. 

	We note that area accessible for observation decreases when Sun 
moves from 
the southern to the northern hemisphere. This is primarily 
due to the functional constraints, providing normal thermal conditions, 
and the possibility of HGCA pointing to the ground tracking antenna.

	As it is shown in this figure, some sources can be observed 
only at the specific time of year and repeated observations are possible 
again in a year. According to FAKERAT, the northern sources can be 
observed within 3-4 months/year, and southern - only within 2 months/year or 
less.

\section{Strategy of Using FAKERAT Package}

	At the present time FAKERAT with graphical interface is working 
on Linux workstations. To start FAKERAT it is necessary 
to run ó Shell script which set environment variables required for 
running libraries of PGPLOT graphical package. An X Window dialog box 
will appear for the user to run FAKERAT software. All entered  parameters 
by means of the graphical interface are saved and transferred to 
the other programs, the copy of which 
is stored in {\bf menu\_defaults.1} file. After starting FAKERAT a user 
has a number of 
possibilities for simulate VLBI ground-space observations. First two 
years of working with FAKERAT showed that the most popular options are 
follows:
(u,v)-coverage as a function of right ascension and declination  the whole 
celestial sphere for a given epoch ({\bf all-sky uvplot}), 
time evolution of (u,v)-coverage 
for a given source, at equally  intervals time ({\bf time-uvplot}), image of 
(u,v)-coverage for given source ({\bf uvplot}), and test constrains 
on date planning 
observations ({\bf constraints}).


\begin{table}
\caption{List of Ground Radio Telescopes and VLBI Networks participating 
in space observations.}
\label{tab:1}
\renewcommand{\arraystretch}{1.}
\begin{tabular}{l|c|r|r|r|r|c}
\multicolumn{7}{l}{ }\\
\hline
GRTs/Networks  & Diam. & \multicolumn{4}{c|}{SEFD} & Time  \\ \cline{3-6}
               & (m)  & P & L & C & K &   (\%) \\
\hline
\multicolumn{7}{c}{EVN$^a$:}\\
\hline
 Arecibo (Ar)                 & 305   & 12   &  3  & 5   &      & 11  \\
 Effelsberg (Eb/Ef)           & 100   & 600  & 19  & 20  &  90  & 51 \\
 Hartebeesthoek (Hh)          & 26    &      & 450 & 795 & 3000 & 6  \\  
 Jodrell Bank (Jb-1)           & 76    & 132  & 65  & 80  &      & 6 \\
 Jodrell Bank (Jb-2)           & 25    &      & 320 & 320 & 910  & 2 \\
 Medicina (Mc)                & 32    &      & 700 & 170 &  700 & 31 \\
 Metsaehovi (Mh)              & 14    &      &     &     & 2608 & 3  \\  
 Nanshan (Urumqi, Ur)         & 25    &      & 300 & 250 & 850  & 5  \\ 
 Noto (Nt)                    & 32    & 980  & 784 & 260 &  800 & 29 \\ 
 Onsala (On-85)               & 25    &      & 320 & 600 &      & 5  \\ 
 Sheshan (Shanghai, Sh)       & 25    &      & 670 & 720 & 1700 & 4  \\ 
 Torun (Tr)                   & 32    &      & 300 & 220 & 500  & 21 \\ 
 Westerbork (Wb)              & 14x25 & 150  & 40  & 60  &      & 26 \\
 Yebes (Ys)                   & 40    &      &     & 160 & 200  & 50  \\  
 Badary (Bd)                  & 32    &      & 330 & 200 & 710  & 20  \\  
 Svetloe (Sv)                 & 32    &      & 360 & 250 & 710  & 27  \\  
 Zelenchuck (Zc)              & 32    &      & 300 & 400 & 710  & 26  \\  
\hline
 Robledo (Ro70, DSS63)       & 70    &      & 35 &     & 83   & 17  \\  
\hline
\multicolumn{7}{c}{LBA$^b$:}\\
\hline
 Narrabri, ATCA (At)         & 1x22  &      & 340 & 350 & 530  & 5  \\  
 Ceduna (Cd)                 & 30    &      &     &     & 2500 & 2  \\  
 Hobart (Ho)                 & 26    &      & 420 &     & 1800 & 5  \\  
 Mopra (Mp)                  & 22    &      & 340 & 350 & 900  & 2  \\  
 Parkes (Pa)                 & 64    &      & 42  & 110 & 810  & 2  \\  
 Tidbinbilla (Ti, DSS43)     & 70    &      & 23  &     & 60   & $<$ 1 \\  
 \hline
\multicolumn{7}{c}{VLBA$^c$:}\\
\hline
 VLBA\_SC, VLBA\_FD,         & 25    &  2227    & 303 & 210 & 502  & $<$ 1  \\  
 VLBA\_PT, VLBA\_OV          &       &          &     &     &      &      \\  
\hline
ASKAP (As)                   & 1x12   &       &  ?   &   &    & 3  \\  
GBT$^d$ (Gb)                 & 100 &  11 &  9   & 10  &  27  & 22  \\  
PT-70 (Ev)                   & 70   &       &  19   & 19  & 110   & 47  \\  
Usuda (Us)                   & 64   &       &  69   & 69  & ?   & 4  \\  
VLA27$^e$ (Y)                & 1x25 &  & 420 & 310 & 560  & $<$ 1  \\  
Warkworth (Ww)               & 12   &       &   ?  &   &    & 3  \\  
Ooty                         &  530mx30m    &    ?   &       &     &       & $<$ 1  \\
 \hline
 {\it KRT} RadioAstron    & 10           & 19000 & 3400 & 10500 & 30000 & 100  \\
\hline
\multicolumn{7}{l}{ }\\
\multicolumn{7}{l}{$^a$http://www.evlbi.org/user\_guide/EVNstatus.txt}\\
\multicolumn{7}{l}{$^b$http://www.atnf.csiro.au/vlbi/documentation/vlbi\_antennas}\\
\multicolumn{7}{l}{$^c$http://science.nrao.edu/facilities/vlba/docs/manuals/oss2013b}\\
\multicolumn{7}{l}{$^d$http://science.nrao.edu/facilities/gbt/proposing/GBTpg.pdf}\\
\multicolumn{7}{l}{$^e$http://science.nrao.edu/facilities/vla/docs/manuals/oss2013a/performance/sensitivity}\\  
\end{tabular}
\end{table}

	Below is list of main actions that should be performed before modeling 
VLBI observation:
\begin{itemize}
\item connect orbit to the FAKERAT package (i.e. in {\bf orbit} directory 
of the  FAKERAT package should be given a symbolic link {\bf ra\_orbit} to 
the chosen orbit; 
\item choose a tracking station for data transfer ({\bf tracking station}: 
PUSCHINO OR GBANK-5); 
\item set the observation date ({\bf obs-year}, {\bf obs-month} and {\bf obs-day}); 
\item set the start time of observations ({\bf star hh:mm:ss}); 
\item set the end time of observations ({\bf stop hh:mm:ss}); 
\item set the receiver's frequency ({\bf observing band}); 
\item set the integration time ($\boldsymbol\tau$ (s)); 
\item specify radio source ({\bf source}); 
\item set the right ascension ({\bf RA hh:mm:ss.ss}) and declination 
({\bf Dec dd:mm:ss.ss}), and 
\item choose ground radio telescopes ({\bf telescopes}).
\end{itemize}
	Before starting the work, a user should 
set the time period when the observation is possible to perform. This task may 
be done using two options: {\bf all-sky uvplot} and {\bf time-uvplot}. 
In case if observation is impossible, 
user can find out the reasons using {\bf constraints} option. 
Typical (u,v)-coverage of grid points for the all 
celestial sphere when the Sun is in the northern part and in the southern part 
of celestial sphere   are presented be displayed in Fig.~\ref{fig:2}. 
In this examples ó-band 
(see below), and GTS in Puschino were used. The list of the ground telescopes 
is given in the upper part of the each figure. In the figure, you can see the 
areas of the celestial sphere that are not possible to observe due 
constraints. The projections of the Galaxy plane (GP) and the SC orbital plane 
onto the celestial sphere are given. Apocenter and pericenter of the SC orbit 
are marked as \begin{picture}(20,20)
\put(10,4){\circle{15}}
\put(5,1){$A$}
\end{picture} and
\begin{picture}(20,20)
\put(10,5){\circle{15}}
\put(5,1){$P$}
\end{picture} respectively.

	Once again, from a comparison of these two figures one can see that, 
when the Sun is in the northern part of celestial sphere, there are significant 
constraints connected with both the Sun and on the capability to point HGCA to 
GTS.

	The ground radio telescopes involved into space-VLBI 
experiments during ESP stage are shown in Table~\ref{tab:1}. Sensitivity of two radio 
telescopes in interferometric mode can be estimated as follows:
\begin{eqnarray}
\sigma_{i,j}=1/\psi\sqrt{SEFD_iSEFD_j/2B\tau} \nonumber 
\end{eqnarray}
where $\psi$ is an efficiency factor with respect to the unquantized case, 
$\tau$ is integration time in sec., $B$ is the receiver bandwidth in Hz, 
$SEFD_{i,j}$ is the system equivalent  flux densities for $i$ and $j$
of the radio telescope in Jy, respectively. For combination of levels 
with one-bit (space radio telescope) and two-bit (ground radio telescope)  
clipped signals  $\xi=0.67$, and for two two-bit 
clipped signals (two ground radio telescopes) $\xi=0.881$. 
We assumed that signal is registered at the Nyquist 
frequency. $SEFD$ values are given in Table~\ref{tab:1}.

\section{Frequency Specification of On-board System of Receivers}

	Receiver complex of SC includes on-board system of four receivers 
that enables to obtain a signal at four wavelengths. Each receiver 
(except C-band) has two independent channels with  left 
(LCP) and right (RCP) circular polarization signals. In C frequency 
ranges can be worked only with one polarization. RadioAstron IF-to-Video 
converter (Formatter) produces 
two 16 MHz bands (USB and LSB) in
frequency ranges L, C, and K. In P-band, receiver contains the signal 
only for upper sub-band.

Each frequency channel can receive a signal at one of the two bands, 
apart from each other at 8~MHz:
\begin{itemize}
\item P band, with central frequencies 308~MHz or 316~MHz and 16~MHz 
bandwidth; 
\item L-band with central frequencies 1660~MHz or 1668~MHz and 60~MHz 
bandwidth;
\item C-band with central frequencies 4828~MHz or 4836~MHz and 110~MHz 
bandwidth and 
\item K-band with central frequencies 22228~MHz or 22236~MHz 
with eight sub-bands for multifrequency synthesis, with four sub-bands 
for observations of narrow spectral lines and 150~MHz bandwidth.
\end{itemize}

Central frequencies of eight sub-bands in K-band for multifrequency synthesis
apart from each other at 960~MHz: 
F$_{-4}$ = 18388~MHz or 18396~MHz, 
F$_{-3}$ = 19348~MHz or 19356~MHz, 
F$_{-2}$ = 20308~MHz or 20316~MHz, 
F$_{-1}$ = 21268~MHz or 21276~MHz, 
F$_0$ = 22228~MHz or 22236~MHz, 
F$_1$ = 23188~MHz or 23196~MHz, 
F$_2$ = 24148~MHz or 24156~MHz and 
F$_3$ = 25108~MHz or 25116~MHz. 

Central frequencies of four sub-bands in K-band for spectral observations
apart from each other at 32~MHz: 
F$_0$ = 22228~MHz or 22236~MHz, 
F$_{0-1}$ = 22196~MHz or 22204~MHz, 
F$_{0-2}$ = 22164~MHz or 22172~MHz and 
F$_{0-3}$ = 22132~MHz or 22140~MHz. 
More information about the 
work of on-board scientific complex a user can find in work \cite{ka}.

	The FAKERAT package provides the ability to calculate of 
(u,v)-coverage for frequency syntheses in K-band using 8 sub-bands 
from 1.19 to 1.63 cm.  Fig.~6f-6k of the work \cite{ka} show examples of 
the corresponding evolution of K-band (u,v)-coverage for two edge sub-bands 
for 2013, 2014 and 2015, for the radio galaxy M87 and Cen~A involving SRT 
and eight ground radio telescopes (Green Bank, Goldstone, Effelsberg,  
Jodrell Bank, Evpatoria, Parkes, Tidbinbilla, and Robledo). Significant 
ellipticity observed in the (u,v)-coverage  can be reduced in the future 
using orbit correction.

\section{Conclusion}

About 600 interferometric sessions of RadioAstron project at all 
four operational wavelengths were planned and successfully observed by the 
end of June 2013. All planning was 
done using FAKERAT package, including the first test observations of the 
Moon. When  Moon observation  was planned, some SC functional constraints in 
the FAKERAT package have been violated. Total time of the scientific 
observations during this period 
 was more than 712 hours.\\[15pt]

\noindent{\it Acknowledgments.}
The RadioAstron project is led by the Astro Space Center of the Lebedev 
Physical Institute of the Russian Academy of Sciences and the Lavochkin 
Scientific and Production Association under a contract with the Russian 
Federal Space Agency, in collaboration with partner organizations in 
Russia and other countries. The author would like to thank Russian and 
foreign colleagues participating in FAKERAT package development: 
A.V.~Alakoz, M.V.~Popov and V.A.~Soglasnov for online  testing package, 
K.V.~Sokolovsky for testing package in automatic mode without using 
graphical interface, Yu.Yu.~Kovalev for developing network version of 
the package, E.B.~Kravchenko and J.M.~Anderson (Max Planck Institute for 
Radio Astronomy, German) for the revision of the archive of the ground 
radio telescopes, P.A.~Voytsik and C.R.~Gwinn (University of California 
Santa Barbara, USA) for conversion of the package into 64-bit version 
and distribution of that version among IBM PC and Apple users, 
V.G.~Promyslov for accommodation of FAKESAT at PC-platform, V.E.~Yakimov 
for placing the package on ASC FIAN website, and T.S.~Fetisova for 
scientific editing of the paper. In particular, the author would like 
to thank D.W.~Murphy (JPL) for its participation in RadioAstron project. 
Due to that participation effective software for planning observations in 
the space-VLBI international interferometric experiment was 
developed within the shortest possible time.


\begin{thebibliography}{99}
\bibitem[Avdeev et.al.~2012]{ak}
Avdeev V.Yu. et al., 2012, Vestn. FGUP NPO im. S.A.Lavochkina, 4
\bibitem[Kardashev et.al.~2013]{ka}
Kardashev N.S. et al. ARep, 2013, 57, 153
\bibitem[Murphy~1991]{mu1}
D.W. Murphy, Simulations of space VLBI, Radio interferometry: 
Theory, techniques, and applications; Proceedings of the 131st IAU 
Colloquium, ASP Conference Series (ASP: San Francisco), 1991, vol. 19, p. 107
\bibitem[Murphy~2006]{mu2}
Murphy~D.W., VSOP-2 Mission Simulations Using the JPL-developed Fakesat 
Software, 36th COSPAR Scientific Assembly. Held 16-23 July in Beijing, 
China, 2006, P.2496 
\end{thebibliography}
\end{document}